# A lightweight design for serverless Function-as-a-Service

Ju Long, Texas State University, Hung-Ying Tai, Shen-Ta Hsieh, and Michael Juntao Yuan, Second State LLC

**Abstract**

FaaS (Function as a Service) allows developers to upload and execute code in the cloud without managing servers. FaaS offerings from leading public cloud providers are based on system microVM or application container technologies such as Firecracker or Docker. In this paper, we demonstrate that lightweight high-level runtimes, such as WebAssembly, could offer performance and scaling advantages over existing solutions, and could enable finely-grained pay-as-you-use business models. We compared widely used performance benchmarks between Docker native and WebAssembly implementations of the same algorithms. We also discuss the barriers for WebAssembly adoption in serverless computing, such as the lack of tooling support.

## Introduction

Serverless computing is an approach to build and deploy cloud services without having to manage servers. Compared with IaaS (Infrastructure-as-a-Service), serverless computing offers a higher level of abstraction, and hence promises to deliver higher developer productivity and lower operation costs. The key characteristics of serverless computing include resource elasticity, zero ops, and pay-as-you-use [1].

The original cloud service, AWS S3 (Simple Storage Service), is serverless, but it is only for storage. However, it proves difficult to scale code execution. IaaS became the dominant cloud service. Developers provision virtual machines (servers) to deploy applications. The developer is responsible for planning, provisioning, deploying, and maintaining the servers. Even with more advanced PaaS (Platform-as-a-Service), developers get virtual servers pre-installed with a software stack, and they must pay based on allocated resources (e.g., CPU and memory) as opposed to actual usage.

In recent years, a new generation of cloud services, pioneered by AWS Lambda, enable developers to create web services based on code functions directly. Developers do not need to provision or manage servers. The cloud provider automatically deploys an execution environment when requests come in, and scales the environment based on demand [2]. That approach is known as Function-as-a-Service (FaaS) [3]. Industry reports indicate that FaaS is already popular among cloud users, and could grow into a dominant form of future cloud computing [4].

The rise of FaaS is enabled by technology advances in virtualization and containerization, which makes it possible to frequently start and stop the execution environment with minimal overhead. Leading public cloud providers use a combination of system VMs and application containers as infrastructure for FaaS. The system VMs are virtual machines that provide isolation at system or hardware level, such as hypervisor-based VMs. They are in contrast with high-level software-based language VMs, such as the Java VM, JavaScript runetime, and WebAssembly VM. Compared with system VMs, application containers like Docker are faster and lighter [5], but less secure [6].

- The AWS Lambda is based on a lightweight virtualization technology called Firecracker [7]. Also known as microVM, Firecracker is lighter and faster than traditional system VMs, and hence is suitable for the FaaS workload.
- IBM Cloud Functions utilizes Docker containers to provide isolation.
- Microsoft Azure Functions is compatible with Docker, but with added VM protection when deployed on the public cloud [8].

- Google Cloud Functions took a compromise approach. Its gVisor engine is designed to safely run Docker in public cloud. However, gVisor is 2x slower than plain Docker [9].

Even with those performance-optimized system VMs and application containers, FaaS still suffer from performance and scalability issues.

First, cold start is slow [10]. To provision and start a microVM or container could take seconds. This is a key issue for FaaS, where each function execution could require a new microVM or container. This problem also lead to inconsistent and unpredictable performance since the function could run faster when an execution environment is "warmed up".

Second, the VM or container must setup a runtime software stack, including operating system-specific standard libraries, for every function call. The result is a large footprint.

Finally, the current generation of VM or container-based FaaS solutions bill users for coarse-grained resource consumption, such as allocated memory and execution time. They do not support fine-grained usage billing, such as CPU cycles for each function execution.

In this paper, we will discuss an emerging FaaS approach that does not depend on VMs or containers.

## A lightweight approach

Serverless FaaS must provide a secure and isolated execution environment for user-submitted and untrusted code. It has been suggested that WebAssembly is a lighter and faster alternative for FaaS [11, 12]. Unlike system VMs and containers, WebAssembly is a high-level language VM runtime.

WebAssembly started as a collaboration to improve JavaScript performance inside web browsers. Since it is designed to execute code downloaded from the web, the WebAssembly runtime provides a Software-based Fault Isolation (SFI) [13] sandbox for untrusted code.

WebAssembly supports multiple programming languages, such as C/C++ and Rust. The WebAssembly System Interface (WASI) standard [14] allows WebAssembly runtimes to interface with operating system resources, which makes it suitable for server-side applications.

WebAssembly functions can run securely and in isolation. Those functions can be started and stopped on-demand across different underlying platforms without any code change. Since WebAssembly provides abstractions at the opcode level, it can precisely measure finely-grained resource consumptions at runtime.

The most important advantages WebAssembly promises over system VMs, microVMs, and containers are performance and footprint. Smaller cloud providers, such as Fastly and Cloudflare, are already offering WebAssembly-based FaaS on their edge networks. In this study, we evaluate performance of leading WebAssembly runtimes against Docker containers. Docker is chosen as the benchmark comparison because it is the most widely used application container and delivers state-of-the-art performance for non-WebAssembly FaaS.

WebAssembly's performance advantages over Docker in FaaS had been documented in the literature [11, 12]. But our study has several noticeable differences. Instead of timing an integrated use case, our study evaluates standard performance benchmarks to eliminate the complexity of the host environment, such as multi-thread efficiency or the disk I/O, which are not directly related to WebAssembly vs Docker runtimes. Furthermore, we evaluate several leading WebAssembly runtimes with different optimization strategies. When possible, we execute WebAssembly functions in the standalone WASI mode as opposed to relying on host runtimes such as Node.js to bootstrap from slow JavaScript.

# Performance evaluation

We conducted experiments on two large cloud platforms, AWS and Microsoft Azure. Both systems run Ubuntu Linux 20.04 TLS. The hardware configurations are as follows.

- AWS: m5.2xlarge instance. Intel CPU 2.30GHz with 4 CPU cores and 32GB of RAM.
- Azure: Standard_D4_v3 instance. Intel CPU 2.50GHz with 4 CPU cores and 16GB of RAM.

The following runtimes are evaluated.

- The latest stable Docker release, v19.03.8. A Docker image for Ubuntu Linux 18.04 was used to start the Docker container.
- V8, v8.1.307 [15], is the WebAssembly runtime developed by Google. It is pre-installed in many web browsers.
- Lucet, v0.6.1 [16], from the Byte Alliance, is a collaboration between Fastly and Mozilla to build a WebAssembly compiler and runtime.
- Second State Virtual Machine (SSVM), v0.6.0 [17], is a WebAssembly runtime optimized for server side workloads.
- WAVM, nightly/2020-05-28 [18], is a collaborative effort led by Andrew Scheidecker. It aims to support WebAssembly outside of the browser.

For each testing machine and runtime combination, we ran the following benchmark tests. We executed each test 50 times, and calculated the average and standard deviations for completion time. The source code and instructions are publicly available on GitHub. [19]

- The nop test starts the runtime and exits. It evaluates cold start performance.
- The cat-sync test opens a local file to read and exits. It evaluates performance in making operating system calls.
- The following 4 benchmarks are from the Computer Languages Benchmarks Game [20], which provides crowd-sourced benchmark programs for over 25 programming languages. The nbody, repeated 50 million times, is an n-body simulation. The fannkuch-redux, repeated 12 times, measures indexed access to an integer sequence. The mandelbrot, repeated 15000 times, is to generate Mandelbrot set portable bitmap file. The binary-tree, repeated 21 times, allocates and deallocates large numbers of binary trees.

All benchmark tests are written in C. For Docker tests, the test source code is compiled to native binary by the LLVM v10.0.0 toolchain. For WebAssembly tests, the source code is compiled into WebAssembly bytecode using the Emscripten SDK v1.39.17.

Figure 1 shows the execution time results from the nop test. WebAssembly cold starts are at least 10x faster than Docker. Among WebAssembly runtimes, SSVM and Lucet are more than 10x faster than the rest. SSVM and Lucet start in less than 1/500 of the time it took Docker to start.

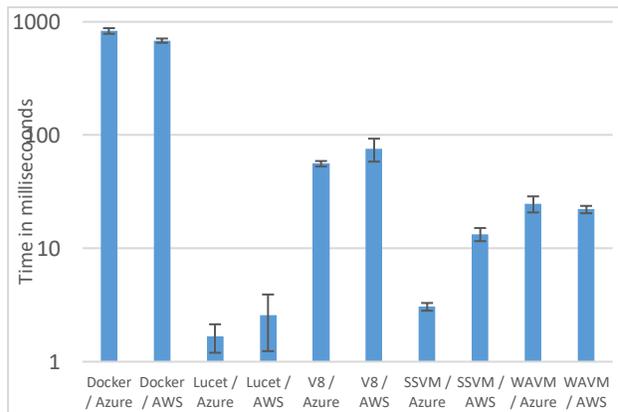

**Figure 1. Cold start time. The smaller the better. The Y-axis is in log10 scale.**

Figure 2 shows the execution time results from the cat-sync test from cold start. With the exception of Lucet, which failed this test, WebAssembly runtimes are at least 10x faster than Docker. Among WebAssembly VMs, SSVM showed the best performance. Docker takes 50x to 150x longer than SSVM to open a file.

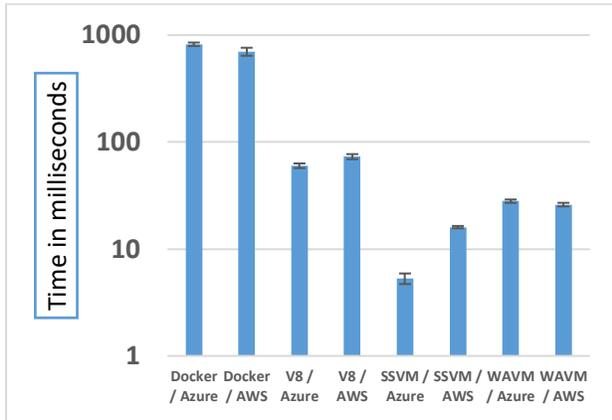

**Figure 2. Time to open and read a file. The smaller the better. The Y-axis is in log10 scale.**

Table 1 shows the execution time results from the 4 computational benchmarks. Since each benchmark test is executed repeatedly in a loop during each test run, those numbers indicate warm start performance. WebAssembly runtimes are about 10% to 50% faster than Docker with the exception of Lucet, which crashed in several tests, and generally performed poorly.

**Table 1. Execution time for computational tests. The smaller the better. All numbers are in seconds.**

|        | binary-tree | fannkuch-redux | mandelbrot | nbody |
|--------|-------------|----------------|------------|-------|
| Azure  |             |                |            |       |
| Docker | 16.9±0.9    | 26.1±0.1       | 16.2±0.9   | 4.1±0.05 |
| Lucet  | Failed      | 56.6±0.1       | Failed     | 4.67±0.03 |
| SSVM   | 12.2±0.1    | 32.8±0.2       | 12.7±0.1   | 3.75±0.03 |
| V8     | 17.4±0.1    | 30.0±0.2       | 10.45±0.04 | 3.38±0.02 |
| WAVM   | 13.4±0.1    | 29.4±0.1       | 11.8±0.07  | 3.74±0.02 |
| AWS    |             |                |            |       |
| Docker | 29.9±0.1    | 39.2±0.1       | 11.29±0.09 | 4.14±0.06 |
| Lucet  | Failed      | 67.8±0.1       | Failed     | 5.52±0.08 |
| SSVM   | 13.4±0.1    | 38±0.1         | 10.8±0.2   | 3.92±0.08 |
| V8     | 19.2±0.2    | 40.±0.8        | 10.6±0.1   | 3.69±0.08 |
| WAVM   | 15.1±0.1    | 36.5±0.1       | 9.96±0.08  | 3.95±0.08 |

Besides the execution time performance advantages, WebAssembly also has a much smaller footprint than Docker. The Docker image for Ubuntu 18.04 is 188MB, while the SSVM is less than 6MB. In general, WebAssembly runtimes are 1/10 of the size of full Docker images for FaaS.

## Challenges and Discussions

While the WebAssembly runtimes outperform Docker in many computing benchmarks, there are also many challenges that could hinder their adoption.

First, programming language support on WebAssembly is limited. While the LLVM toolchain supports generating WebAssembly bytecode from 20+ languages, only a handful of languages are well supported in practice. Currently, C/C++, Rust, and AssemblyScript are the best supported programming languages for WebAssembly. That means developers need to write their functions in those languages to take advantage of WebAssembly-based FaaS.

Second, compared with containers and microVMs, WebAssembly has a different model to access the operating system. System features, such as networking and multithread management, are just becoming available to WebAssembly. However, WebAssembly access to those resources are governed by high-level security policies (i.e., capability-based security), and could be both performant and secure. In this study, we did not evaluate I/O performance of WebAssembly runtimes beyond simple file operations. Multithreaded I/O is an important aspect of real world applications, and it is a topic of future research as related WebAssembly specifications mature.

Finally, there is currently no industry standard management and orchestration tools for WebAssembly runtimes. Most use home grown solutions to start and stop WebAssembly runtimes on-demand. We anticipate such tools will become available as WebAssembly adoption increases.

# Conclusion

We evaluated the use of WebAssembly runtimes in serverless FaaS. Compared with containers or microVMs, WebAssembly could enable faster and more scalable services with fine-grained pay-as-you-use billing.

WebAssembly tooling and runtimes optimized for cloud servers, are the main hurdles the industry must overcome to see wide adoption of this technology. Developers probably need to rewrite part of their applications in C/C++ or Rust in order to take advantage of WebAssembly FaaS today.

## References

[1] Castro et al. "The Rise of Serverless Computing," COMMUNICATIONS OF THE ACM, vol 62, No. 12, p44, 2019
[2] Rajan, A. "Serverless Architecture - A Revolution in Cloud Computing," 2018 Tenth International Conference on Advanced Computing. DOI: 10.1109/ICoAC44903.2018.8939081, 2018
[3] Castro et al. "Serverless Programming," 2017 IEEE 37th International Conference on Distributed Computing Systems. p2658, 2017
[4] Jonas et al. "Cloud Programming Simplified: A Berkeley View on Serverless Computing", arXiv:1902.03383, 2019
[5] Zhang et al. "A Comparative Study of Containers and Virtual Machines in Big Data Environment," IEEE 11th International Conference on Cloud Computing. p178, 2018
[6] Di Pietro, R. "To Docker or Not to Docker: A Security Perspective," IEEE Cloud Computing 3(5):54-62. 2016
[7] Agache et al. "Firecracker: Lightweight Virtualization for Serverless Applications," 17th USENIX Symposium on Networked Systems Design and Implementation. p419, 2020
[8] Wang et al. "Peeking Behind the Curtains of Serverless Platforms," Proceedings of the 2018 USENIX Conference. p133, 2018
[9] Young et al. "The True Cost of Containing: A gVisor Case Study," 11th USENIX Workshop on Hot Topics in Cloud Computing, 2019
[10] Lloyd et al. "Serverless Computing: An Investigation of Factors Influencing Microservice Performance," IEEE International Conference on Cloud Engineering. p159, 2018
[11] Hall A. & Ramachandran, U. "An Execution Model for Serverless Functions at the Edge," International Conference on Internet-of Things Design and Implementation. 2019.
[12] Shillaker, S. & Pietzuch P. "Faasm: Lightweight Isolation for Efficient Stateful Serverless Computing," arXiv:2002.09344
[13] Wahbe, R et al. "Efficient Software-based Fault Isolation," ACM Symposium on Operating Systems Principles (SOSP), 1993.
[14] WebAssembly Community Group. "WebAssembly System Interface," Available: https://github.com/WebAssembly/WASI, 2020
[15] Google, "The V8 JavaScript Engine," Available: https://github.com/v8/v8, 2020
[16] Byte Alliance, "Lucet, the sandboxing WebAssembly compiler,", Available: https://github.com/bytecodealliance/lucet, 2020
[17] Second State, "The Second State WebAssembly Virtual Machine," Available: https://github.com/second-state/SSVM, 2020
[18] WAVM, "WAVM is a WebAssembly virtual machine, designed for use in non-web applications," Available: https://wavm.github.io/, 2020
[19] Second State, "wasm32-wasi benchmarks," Available: https://github.com/second-state/wasm32-wasi-benchmark, 2020
[20] Gouy, I. "The Computer Language Benchmarks Game," Available: https://benchmarksgame-team.pages.debian.net/benchmarksgame/, 2020